%% file: main.tex
\documentclass[sigconf]{acmart}

\usepackage{booktabs} 

\usepackage{algorithm}
\usepackage{algpseudocode}

\usepackage{listings}
\usepackage{balance}

\usepackage{graphicx, subcaption}
\usepackage{colortbl} 
\usepackage{multirow}
\usepackage{multicol}
\usepackage{caption}
\usepackage[utf8]{inputenc}
\usepackage[T1]{fontenc}
\usepackage{enumitem}
\usepackage{mathtools}
\usepackage{amsmath}
\usepackage{hyperref}
\usepackage{amsfonts}
\usepackage{color}
\usepackage{xcolor}

\usepackage{array, booktabs, ragged2e}

\setcopyright{acmcopyright}

\copyrightyear{2025}
\acmYear{2025}
\setcopyright{cc}
\setcctype{CC-BY-NC-SA}
\acmConference[SAC '25]{The 40th ACM/SIGAPP Symposium on Applied Computing}{March 31-April 4, 2025}{Catania, Italy}
\acmBooktitle{The 40th ACM/SIGAPP Symposium on Applied Computing (SAC '25), March 31-April 4, 2025, Catania, Italy}\acmDOI{10.1145/3672608.3707950}
\acmISBN{979-8-4007-0629-5/25/03}

\begin{document}
\title{Deciphering Social Behaviour: a Novel Biological Approach For Social Users Classification}
  
\renewcommand{\shorttitle}{A Novel Biological Approach For Social Users Classification}

\author{Edoardo Allegrini}
\orcid{0009-0003-8842-6873}
\affiliation{%
  \institution{Computer Science Department, Sapienza University of Rome, Italy} 
}
\email{allegrini@di.uniroma1.it}

\author{Edoardo Di Paolo}
\orcid{0000-0001-9216-8430}
\affiliation{%
  \institution{Computer Science Department, Sapienza University of Rome, Italy} 
}
\email{dipaolo@di.uniroma1.it}

\author{Marinella Petrocchi}
\orcid{0000-0003-0591-877X}
\affiliation{%
  \institution{Istituto di Informatica e Telematica, CNR}
  \city{Pisa} 
  \country{Italy}
  }
\affiliation{%
  \institution{Scuola IMT Alti Studi Lucca}
  }
\email{marinella.petrocchi@iit.cnr.it}

\author{Angelo Spognardi}
\orcid{0000-0001-6935-0701}
\affiliation{%
  \institution{Computer Science Department, Sapienza University of Rome, Italy} 
}  
\email{spognardi@di.uniroma1.it}

\renewcommand{\shortauthors}{E. Allegrini et al.}

\begin{abstract}
Social media platforms continue to struggle with the growing presence of social bots—automated accounts that can influence public opinion and facilitate the spread of disinformation. Over time, these social bots have advanced significantly, making them increasingly difficult to distinguish from genuine users. Recently, new groups of bots have emerged, utilizing Large Language Models to generate content for posting, further complicating detection efforts.
 This paper proposes a novel approach that uses  algorithms to measure the similarity between DNA strings, commonly used in biological contexts, to classify social users as bots or not. 
 Our approach begins by clustering social media users into distinct macro species based on the similarities (and differences) observed in their timelines. These macro species are subsequently classified as either bots or genuine users, using a novel metric we developed that evaluates their behavioral characteristics 
 in a way that mirrors biological comparison methods. 
This study extends beyond past approaches that focus solely on identical behaviors via analyses of the accounts' timelines. By incorporating new metrics, our approach systematically classifies non-trivial accounts into appropriate categories, effectively peeling back layers to reveal non-obvious species. 
\end{abstract}

%
%



\maketitle

\input{acm_sac2025/body}

\begin{acks}
This work is partially supported by project SERICS (PE00000014) under the NRRP MUR program funded by the EU - NGEU; by project re-DESIRE (DissEmination of ScIentific REsults 2.0), funded by IIT-CNR; by project `Prebunking: predicting and mitigating coordinated inauthentic behaviors in social media`, funded by Sapienza University of Rome.
\end{acks}

\bibliographystyle{ACM-Reference-Format}
\balance
\bibliography{bibliography} 

\end{document}

%% file: acm_sac2025/body.tex
\section{Introduction}
    \input{latex_template/paper/introduction}

    \input{latex_template/paper/evaluation_results/datasets}

\section{Methodology}

    \autoref{fig:classificationProcess} shows the outline of the procedure we propose for classifying social users as genuine or spambots. First, we preprocess users by encoding their online behavior using character sequences known in the literature as digital DNA sequences, see \autoref{subsec:dna_encoding}. Then we cluster the users into macro species (\autoref{subsec:initial_clustering}) and identify two primary groups (\autoref{subsec:earl_arrang}),  one comprising species that exhibit bot-like behavior, and the other comprising species that exhibit clear real-user behavior.  For the remaining species, whose users are not immediately classifiable as bots or not, we have developed DNA similarity metrics that assign species to one of the two classes, as described in \autoref{subsec:geneticsim}.

    \input{latex_template/paper/methodology/procedure_scheme}

    \input{latex_template/paper/methodology/encoding_DNA}

    \input{latex_template/paper/methodology/users_species}

    \input{latex_template/paper/methodology/early_Gspambot_Ggenuine}

    \input{latex_template/paper/methodology/final_classification_genetic_similarity}

\section{Evaluation and Results}

    \input{latex_template/paper/evaluation_results/results}

    \section{Related work}
    \input{latex_template/paper/related_work}

\section{Conclusions}

\input{latex_template/paper/conclusion}

%% file: latex_template/paper/introduction.tex
The digital era has led to an unprecedented surge in social platform accounts, creating a multifaceted and intricate ecosystem. Among these, automated accounts, often referred to as bots \cite{ferrara2016rise, cresci2020decade}, are particularly noteworthy. These digital entities have garnered significant attention not only due to their widespread presence but also because of their frequent use in spreading misinformation and propaganda \cite{shao2018spread}.
Research into bots and their impacts has evolved into an interdisciplinary field, drawing insights from sociology, computer science, political science, and various other domains. A key research avenue emerged with the understanding that bots, programmed to achieve specific objectives, can operate in a coordinated manner and exhibit similar behavioral patterns.
One prominent modeling and detection technique, based on digital DNA and social fingerprinting \cite{DBLP:journals/expert/CresciPPST16, dna_pres}, has proven especially significant. Digital DNA represents a sequence of characters, each corresponding to a particular action taken by an account, thereby capturing the account's activity timeline.
The associated detection technique, the Social Fingerprinting, uses tools known from bioinformatics to measure the similarity between strings: the intuition is that in a group of accounts, the greater the length of the substring identical for all the accounts, the greater the probability that the group consists of bot accounts. 
The digital DNA modeling approach and the Social Fingerprinting methodology allow to apply DNA analysis techniques that have been available in bioinformatics for decades. The model and detection technique are also platform and technology agnostic, paving the way for various behavioral characterization activities.

Beginning with the assumption that accounts of the same type exhibit similar, if not identical, behaviors, this study aims to thoroughly investigate the nature of accounts that cannot be categorized through precise timeline matching. Using a real-world dataset that includes both genuine users and bots, we first identify two subgroups characterized by distinct bot-like and non-bot-like traits. Then, using DNA similarity algorithms, we also classify those accounts that do not have these distinct traits. Our experiments are conducted using two datasets. The first, known as \textit{Cresci-17} \cite{Cresci_2017}, was originally published in 2017 and has since been widely utilized in numerous studies. The second dataset, referred to as \textit{fox-8}, was recently introduced by Yang and Menczer \cite{fox8} and features bots that use Large Language Models (LLMs) to generate their posts.

\textbf{Contributions.} This research advances current methods of timeline-based behavioral analysis for account classification. Unlike previous approaches that classify accounts as bots based on \textit{identical} behavior patterns across most or all of their timelines, our method also considers those accounts whose behavior is more difficult to categorize. Using new metrics, we systematically cluster the ambiguous accounts and assign each to the appropriate category, distinguishing them as either bots or genuine users.

\textbf{Results.} Our approach demonstrates high effectiveness, achieving an F1 score of 0.96 on the \textit{Cresci-17} dataset. For the \textit{fox-8} dataset-a relatively new and underexplored dataset for which the highest known F1 score is 0.84-our method achieves a comparable F1 score of 0.83.
We emphasize that our approach does not rely on machine learning or deep learning techniques. While this may seem unconventional compared to current trends and may raise concerns about performance, our results suggest otherwise. On a relatively unexplored dataset such as the one containing LLM-driven bots, we achieve an F1 score only 0.01 points lower than the best-known result in the literature, while using a method that remains highly interpretable.

%% file: latex_template/paper/evaluation_results/datasets.tex
\section{Datasets}
In this section, we outline the datasets utilized for testing the proposed approach.

\textbf{Cresci--17.}\label{subsec:c17}
The dataset commonly referred to as \textit{Cresci-17} \cite{Cresci_2017}, named after one of the authors and the year of the paper's publication, is a well-known dataset of genuine and social bot accounts in the literature. This study uses a subset of the original dataset consisting of about 3500 \textit{genuine accounts} (\textbf{gen}) identified through a hybrid crowdsensing approach \cite{hybrid-crowdsensing}, where Twitter users meaningfully responded to simple questions. The subset also includes about 1000 bot accounts, \textit{Social spambots \#1} (\textbf{ss1}), supporting a candidate during the 2014 mayoral campaign in Rome; about 3400 bots, \textit{Social spambots \#2} (\textbf{ss2}), promoting a mobile app for artists; and about 450 bots, \textit{Social spambots \#3} (\textbf{ss3}), spamming Amazon products.

\textbf{Fox--8.}\label{subsec:fox8}
The \textit{fox-8} dataset consists of a botnet of approximately 1000 Twitter/X accounts whose posts are generated by an LLM (likely ChatGPT) and whose content promotes suspicious cryptocurrency and blockchain websites. 
The authors of~\cite{fox8} identified the LLM-powered bots because the latter have accidentally revealed themselves in their tweets. In fact, to mitigate the generation of unwanted content, proprietary LLMs often include safeguards such as responding to user requests with standardized messages that clearly state their identity as AI language models and their inability to comply with the requests. After searching for tweets containing such messages ~\cite{fox8} identified 1140 LLM-powered bots. 
Additionally, 1140 genuine accounts were selected from 4 publicly available datasets \textit{botometer-feedback}, \textit{gilani-17}, \textit{midterm-2018}, and \textit{varol-icwsm} to balance bots and genuine accounts. In summary, \textit{fox-8} consists of 1140 LLM-powered bots and 1140 genuine accounts and the dataset is available at the Bot Repository web page maintained at Indiana University\footnote{\url{https://botometer.osome.iu.edu/bot-repository/}}.

%% file: latex_template/paper/methodology/procedure_scheme.tex
\begin{figure*}[htb]
    \centering
    \hfill
    \begin{subfigure}{.25\textwidth}
        \includegraphics[width=\textwidth]{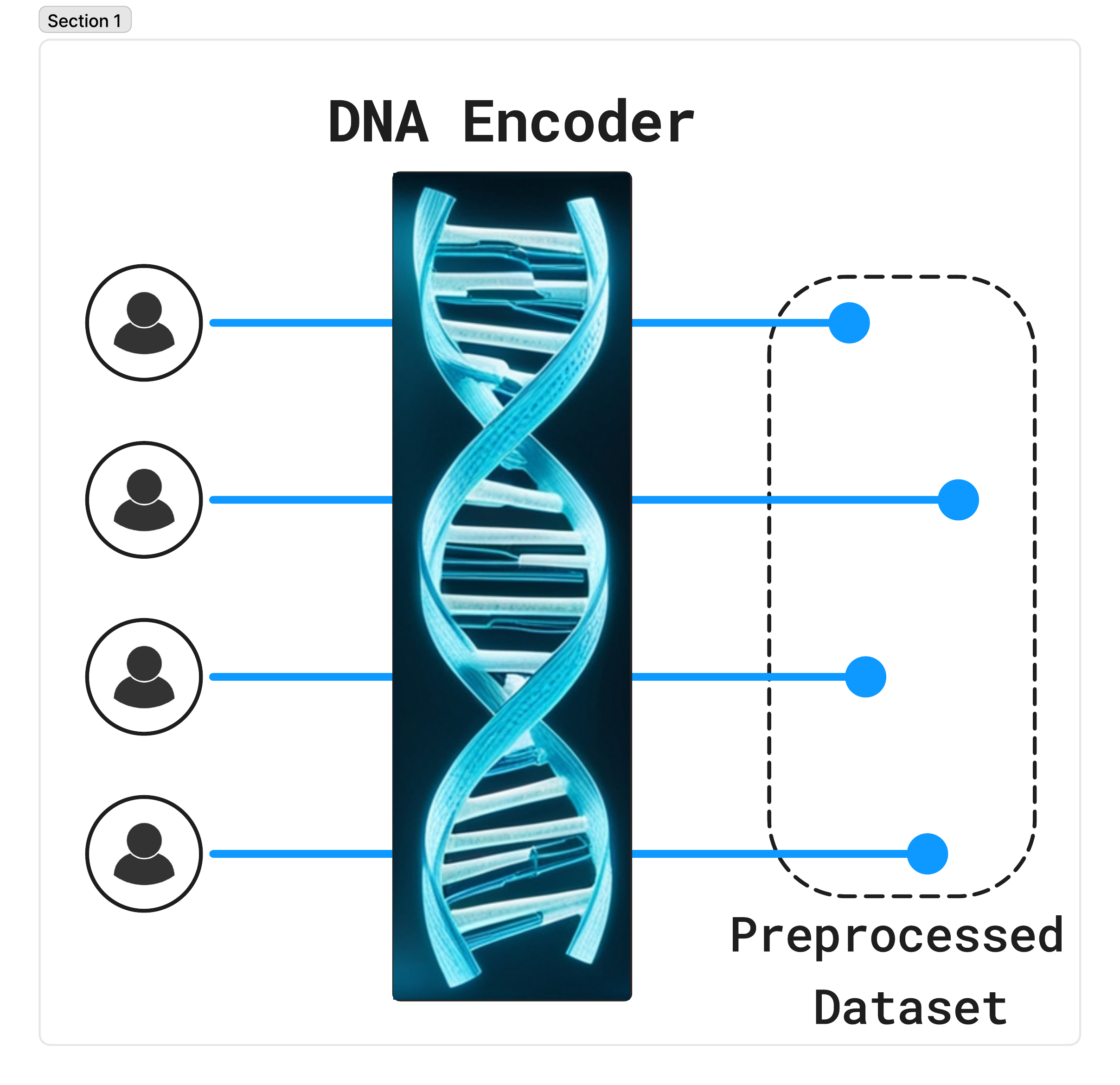}
        \subcaption{Digital DNA (\autoref{subsec:dna_encoding})}
    \end{subfigure}
    \hfill
    \begin{subfigure}{.45\textwidth}
        \includegraphics[width=\textwidth]{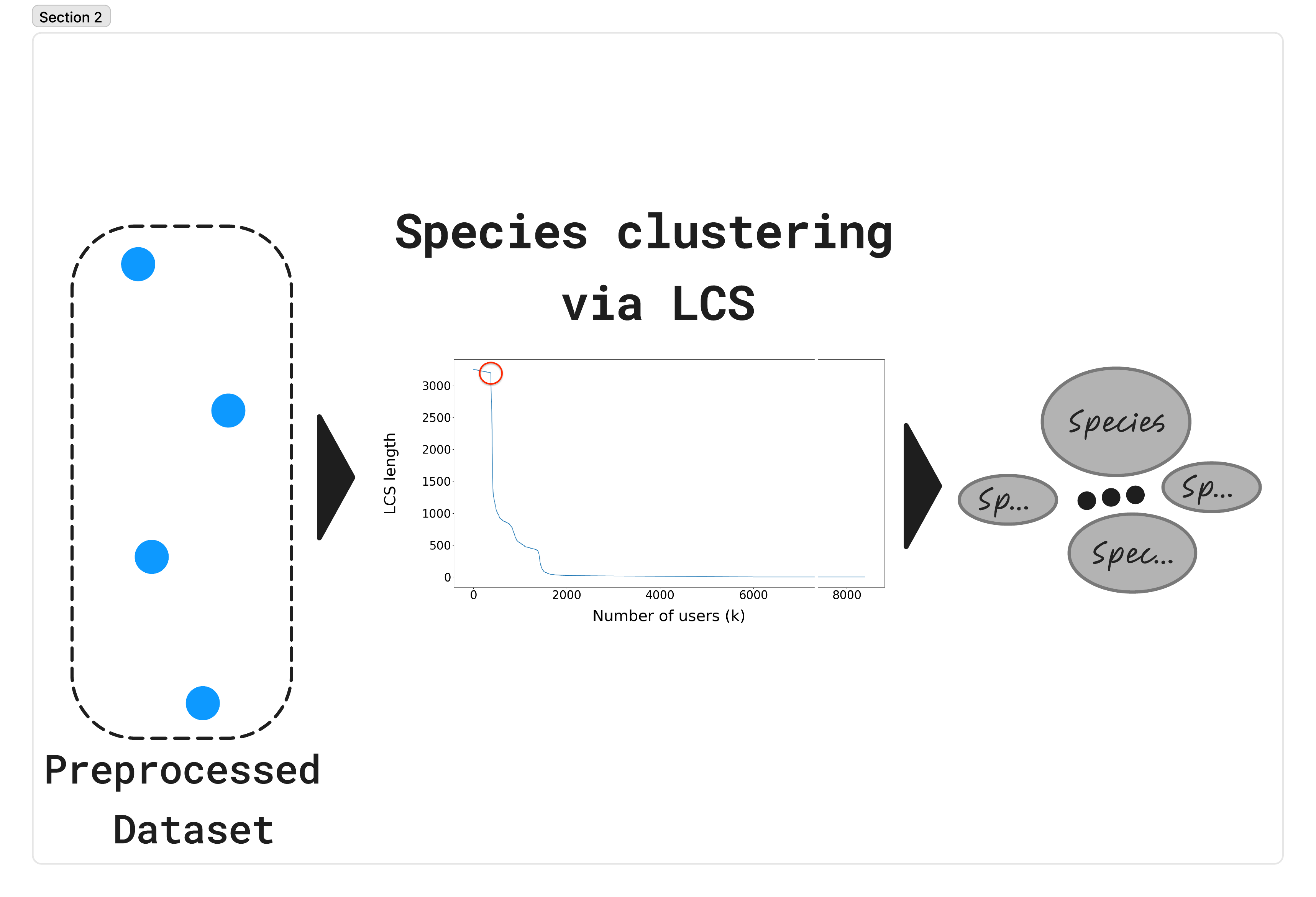}
        \subcaption{Clustering users into species (\autoref{subsec:initial_clustering})}
    \end{subfigure}

    \hfill
    \begin{subfigure}{.25\textwidth}
        \includegraphics[width=\textwidth]{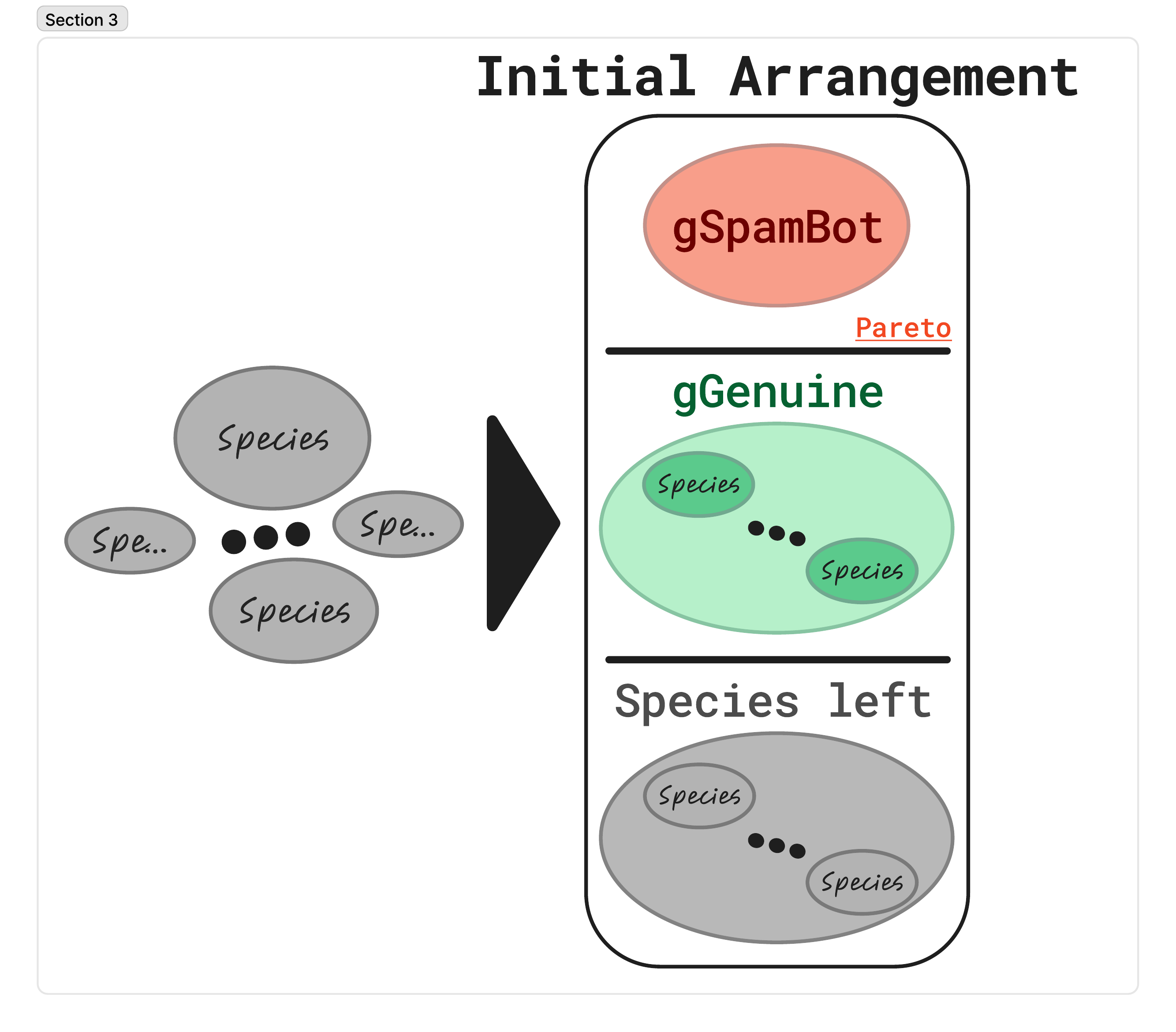}
        \subcaption{Spambot and genuine accounts: initial arrangement (\autoref{subsec:earl_arrang})}
    \end{subfigure}
    \hfill
    \begin{subfigure}{.45\textwidth}
        \includegraphics[width=\textwidth]{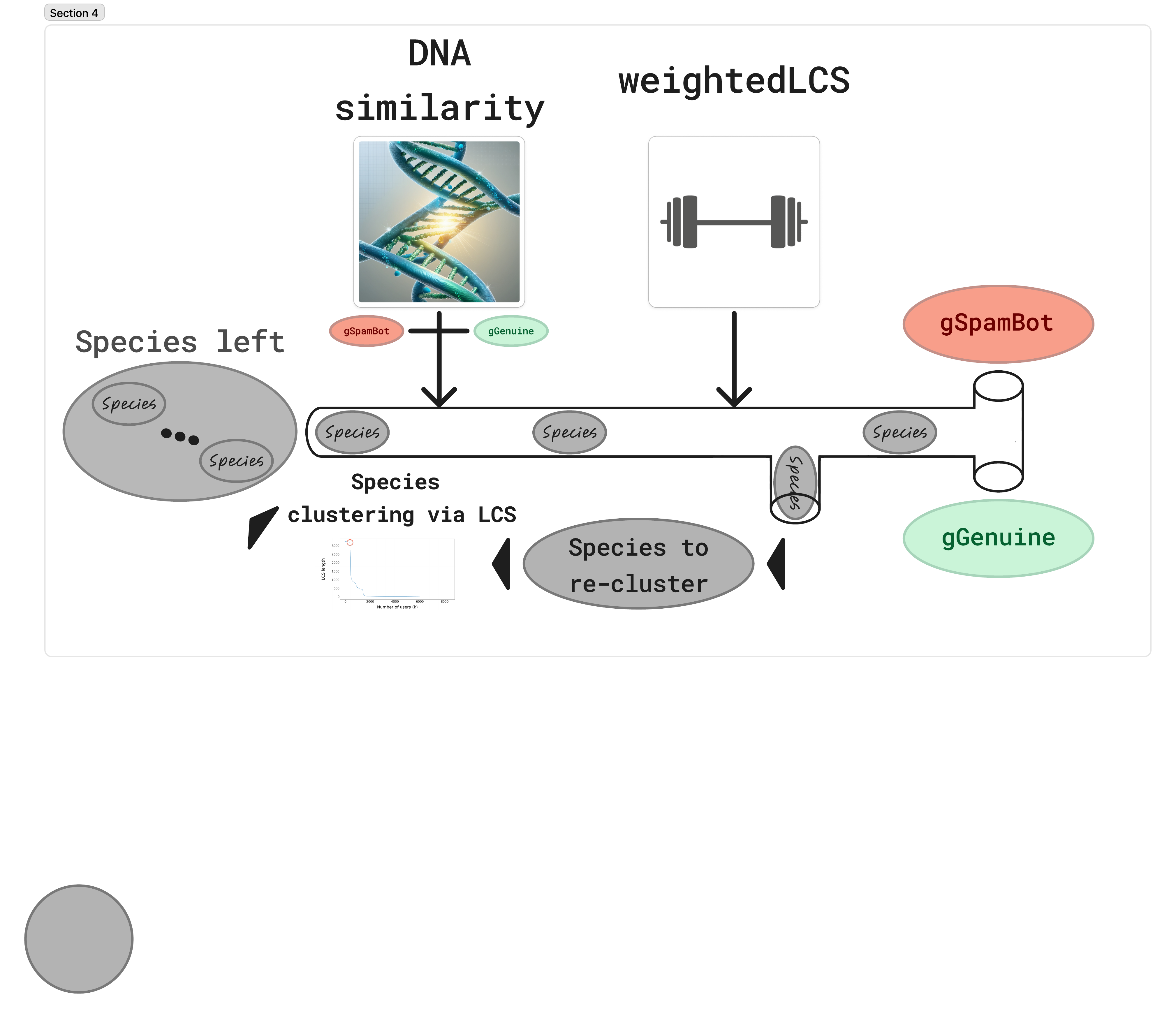}
        \subcaption{Classification of species using DNA similarity (\autoref{subsec:geneticsim})}
    \end{subfigure}
    \hfill    
    \caption{Scheme of the procedure for social users classification\label{fig:classificationProcess}}    
\end{figure*}

%% file: latex_template/paper/methodology/encoding_DNA.tex
\subsection{Digital DNA}\label{subsec:dna_encoding}
    Digital DNA~\cite{dna_pres,DBLP:journals/expert/CresciPPST16} is inspired by biological DNA. Using digital DNA, it is possible to encode the behavioral patterns of a social media account through a condensed sequence of characters derived from an alphabet $\mathbb{B}$, such as the one shown below made up of three characters:\begin{equation}
        \label{eq:alphATC}
        \mathbb{B}^{3}_{\textit{type}} = \begin{Bmatrix}
        A \rightarrow \text{plain tweet}\\ 
        T \rightarrow \text{retweet} \\ 
        C \rightarrow \text{reply}
        \end{Bmatrix}
    \end{equation}

    In \autoref{eq:alphATC}, the alphabet focuses on the type of a tweet. In the preprocessing phase, we encode the user timeline as a sequence defined as:\begin{equation}
        s = (\sigma_1, \sigma_2, ..., \sigma_N), \; \sigma_i \in \mathbb{B}^3_\textit{type} \forall i = 1, ..., N
    \end{equation}

    Alphabets can be enriched with more precise details, such as characterizing users' posts according to their content (an image, a link, a hashtag...).

%% file: latex_template/paper/methodology/users_species.tex
\subsection{Clustering users into species}\label{subsec:initial_clustering}
   Each timeline is encoded with the corresponding string of digital DNA. We now group users into species based on the concept of the Longest Common Substring (LCS) in a set of $k$ accounts \cite{lcs2011}. 
    Given a dataset of $N$ users along with their corresponding digital DNA sequences, we compute the LCS between $k$ users, where $k \in {2,...,N}$, in linear time \cite{lcs2011}. This procedure is based on identifying the LCS among different group sizes of users, ranging from pairs to larger ensembles, denoted by the variable $k$.
    The results are represented in a vector $\mathbb V$, where each element corresponds to the LCS for a given number $k$ of users. 
    
    To provide more detail on how LCS curves are constructed, consider a dataset of four users with DNAs: \{\textit{user1}: \verb|AATTCCA|, \textit{user2}: \verb|TTCAAA|, \textit{user3}: \verb|TTTTTTT|, \textit{user4}: \verb|CCCCCCCT|\}.  Vector $\mathbb V$ is:
    $
        [\verb|TTC|, \verb|TT|, \verb|T|]
    $.
    In fact, the LCS between two users is between \textit{user1} and \textit{user2} (\verb|TTC|), the LCS between three users is between \textit{user1}, \textit{user2} and \textit{user3} (\verb|TT|), and finally the LCS between all four users is \verb|T|. Since the vector $\mathbb V$ contains sub-DNAs belonging to groups of 2, 3, \ldots, N users, we can say whether or not the groups contain $k$ users with similar behavior by analyzing the series of values in $\mathbb V$.
    In particular, as long as the sub-DNAs in $\mathbb V$ remain approximately constant in length, we can infer that users with these sequences exhibit similar behavior. However, if there is a significant drop in the sequence of values in $\mathbb V$ (i.e., $|v_{i+1}|\ll |v_i|$), this indicates that the behavior of the newly added users is significantly different from that of the previous group.
    
    \textbf{Identification of significant drops in $\mathbb V$.}\label{par:first_sign_drop} 
        We present an algorithm for identifying significant drops in the LCS curve. The algorithm is based on basic concepts such as relative change in length (the amount of change over time from an original value) and standard deviation (how widely the data are spread around the mean). 
        From the vector $\mathbb V$, which contains $n$ sub-DNAs, we compute the vector $\mathbb R$, of the ongoing relative change in lengths, such as
        \begin{align}
            \mathbb R = [\frac{|v_2| - |v_1|}{|v_1|}, \frac{|v_3| - |v_2|}{|v_2|}, ..., \frac{|v_n| - |v_{n-1}|}{|v_{n-1}|}]\text{, where } v_i \in \mathbb V
        \end{align}
        Then, we use the standard deviation $\sigma_{\mathbb R}$ of $ \mathbb R$ and a parameter $\tau$ to set a threshold $t$, below which we consider the first significant drop in the LCS curve\footnote{Since the standard deviation is positive and the values in $\mathbb R$ are negative, we have included the minus sign to calculate the threshold.}:
        \begin{align}\label{eq:thresh}
            t = -\tau \times \sigma_\mathbb R
        \end{align}
    Given the values of the vector $\mathbb R$, the first element $r_i \in \mathbb R$ that falls below $t$ will be the first significant drop of the curve and will refer to $v_{i-1} \in \mathbb V$.
    \autoref{tab:thr_eval} shows the performance in terms of classification accuracy considering different values of $\tau$ and different subgroups of the \textit{Cresci-17} and \textit{fox-8} datasets. As we can see, the value of $\tau$ equal to 2 is the one that gives the best performance in terms of accuracy. Therefore, in the following we will perform our experiments with $\tau$ equal to 2. Using \autoref{eq:thresh}, we identify the point where $\mathbb V$ exhibits the first \textit{significant}\footnote{We require $i>20$ to ensure the initial species to have at least 20 users} drop.
    Next, we remove all the users associated with $v_{i-1}$ from the initial dataset and cluster them into a new group called \textit{species}.  Then, we recompute the vector $\mathbb V$ based on the remaining users and repeat the process.
Iterating the above method, we progressively partition the users into several species.

%% file: latex_template/paper/methodology/early_Gspambot_Ggenuine.tex
\subsection{Spambot and genuine accounts: initial arrangement}\label{subsec:earl_arrang}
    In this phase, we identify two key groups among the constructed species (\textit{gSpamBot} and \textit{gGenuine}) looking for users with clear bot and genuine characteristics. The two key groups will guide the subsequent classification of the other species, by including the users belonging to the unclassified species using DNA similarity algorithms (see \autoref{subsec:geneticsim}). The concept that will guide the initial formation of \textit{gSpamBot} and \textit{gGenuine} is the following: the LCS of a species represents the users within it; thus, if the LCS is long, it indicates that the users share similar social behaviors, potentially indicating the presence of social bots. Conversely, a shorter LCS indicates greater variation in social behavior among group members, indicative of uncoordinated action and therefore attributable to genuine accounts.

    \input{latex_template/tables/results_table}

    \subsubsection{Initial gSpambot.}\label{subsubsec:igspambot}
        The formation of the first group of \textit{gSpamBot} was inspired by the Pareto Principle, which seeks to determine the subset of individuals who have the greatest influence on the community as a whole\footnote{Vilfredo Pareto noticed that 80\% of the wealth in Italy was in the hands of 20\% of Italians. This insight later became the eponymous principle that for many phenomena, 20\% of the causes produce 80\% of the effects \cite{pareto1896cours}.}. The process is as follows: first, we sort the species in descending order based on the number of their users, and we consider the top 20\% of the ordered species, namely the $p$ largest ones. The rationale is to consider as the initial \textit{gSpambot} those user groups that not only have the longest LCS, but are also the most numerous. Thus, we adopt the following \textit{weight function} 
        \begin{align}
            \mathbb W(s) = |LCS_s| \times |s|
        \end{align}
        where, for a species $s$, $|LCS_s|$ is its LCS length and $|s|$ is the number of users, and we insert in the \textit{gSpambot} group those species maximizing $\mathbb W(s)$ among the $p$ considered species. Multiple species can reach the same maximum value of $\mathbb W(s)$ because it depends on both the length of the LCS and the number of users, so different combinations can result in the same maximum weight.

    \subsubsection{Initial gGenuine.}\label{subsubsec:iggenuine}

       The initial \textit{gGenuine} is composed of all species whose LCS length is equal to the shortest LCS length found in the species set.
       Including species with a relatively short LCS in \textit{gGenuine} means including users who exhibit behaviors that  do not follow a specific pattern.

%% file: latex_template/tables/results_table.tex
\begin{table}[tbh]
\centering
\caption{Percentage of Accuracy achieved on datasets as the parameter $\tau$ varies.}
\label{tab:thr_eval}
\begin{tabular}{cccccc}
\toprule
$\boldsymbol \tau$ & \begin{tabular}[c]{@{}c@{}}\textbf{gen} $\cup$ \textbf{ss1}\end{tabular} & \begin{tabular}[c]{@{}c@{}}\textbf{gen} $\cup$ \textbf{ss2}\end{tabular} & \begin{tabular}[c]{@{}c@{}}\textbf{gen} $\cup$ \textbf{ss3}\end{tabular} & \textbf{Cresci-17} & \textbf{fox-8} \\ 
\midrule
1 & 99 & \textbf{96} & 99 & 95 & 79 \\ 
1.5 & 99 & 94 & 99 & 95 & 70 \\ 
2 & \textbf{99} & 95 & \textbf{99} & \textbf{95} & \textbf{80} \\ 
2.5 & 99 & 90 & 99 & 93 & 79 \\
3 & 99 & 89 & 99 & 91 & 51 \\ 
3.5 & 24 & 90 & 70 & 92 & 60 \\ 
\bottomrule
\end{tabular}
\end{table}

%% file: latex_template/paper/methodology/final_classification_genetic_similarity.tex
\subsection{Classification of species using DNA similarity}\label{subsec:geneticsim}

    Starting from the two key groups, we use a DNA similarity metric to classify the remaining, unlabeled species.
    A well-established similarity metric is the Levenshtein distance \cite{Levenshtein1965BinaryCC}, which measures the similarity between two strings by calculating the minimum number of single-character changes --insertions, deletions, or substitutions-- required to transform the first string into the other. 
    Moreover, we implemented the \textit{sequence alignment} algorithm, commonly used in biological contexts, to have also a more sophisticated and flexible metric since the Levenshtein distance treats each type of change equally (assigning the exact weight of 1 to insertions, deletions, and substitutions, while a match weights 0). This approach allows us to capture more nuanced similarities between sequences, effectively improving our ability to classify unlabeled species. Since sequence alignment is critical in comparing DNA, RNA, or protein sequences in bio-genetics, we use it to compare the LCS of the unlabeled species and the two key groups. 

    \subsubsection{DNA sequence pairwise alignment.}
\begin{figure*}[htb]
            \centering
            \begin{subfigure}{.45\textwidth}
                \includegraphics[width=\textwidth]{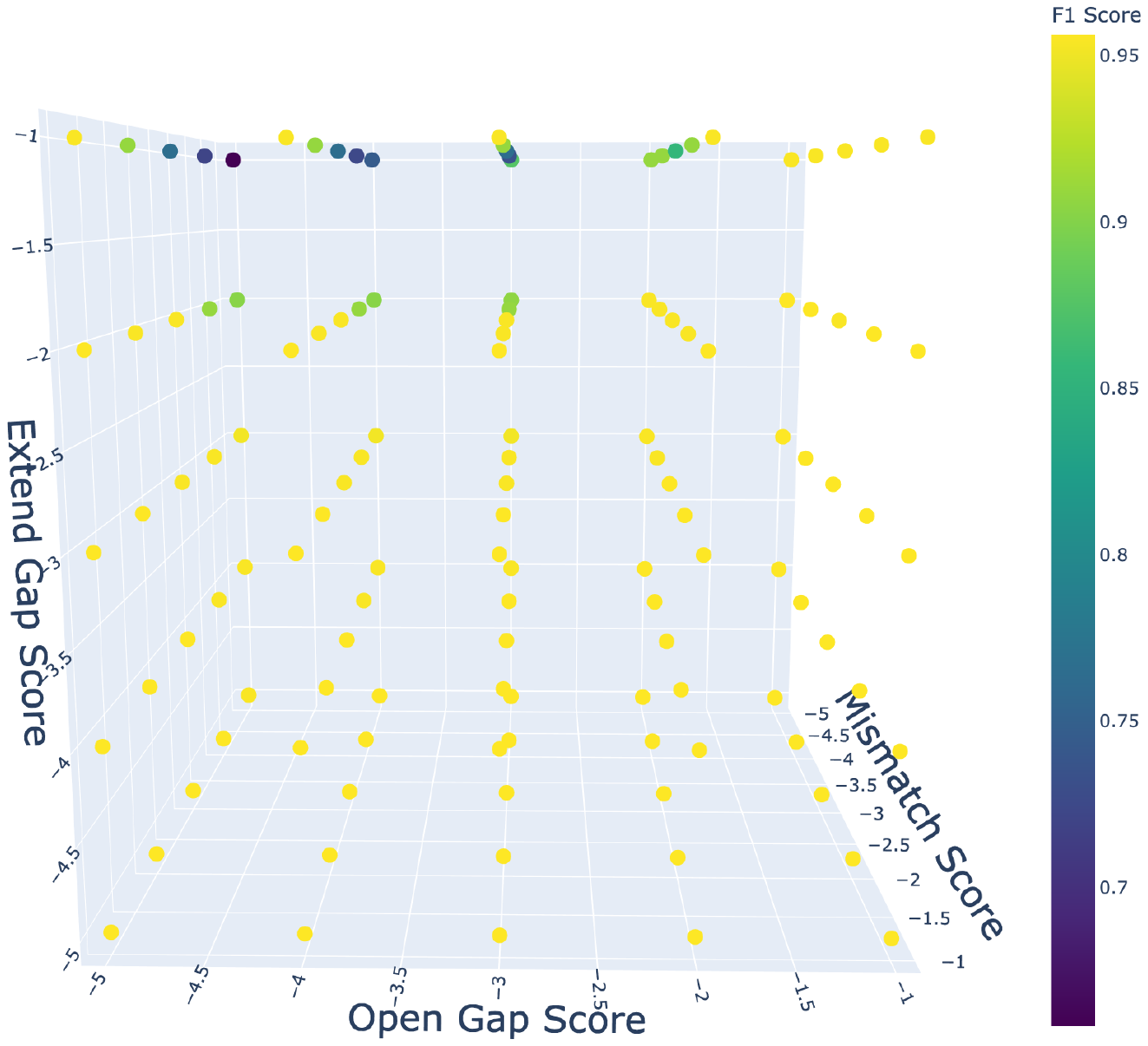}
                \subcaption{\textit{Cresci-17}}
            \end{subfigure}
            \hfill
            \begin{subfigure}{.45\textwidth}
                \includegraphics[width=\textwidth]{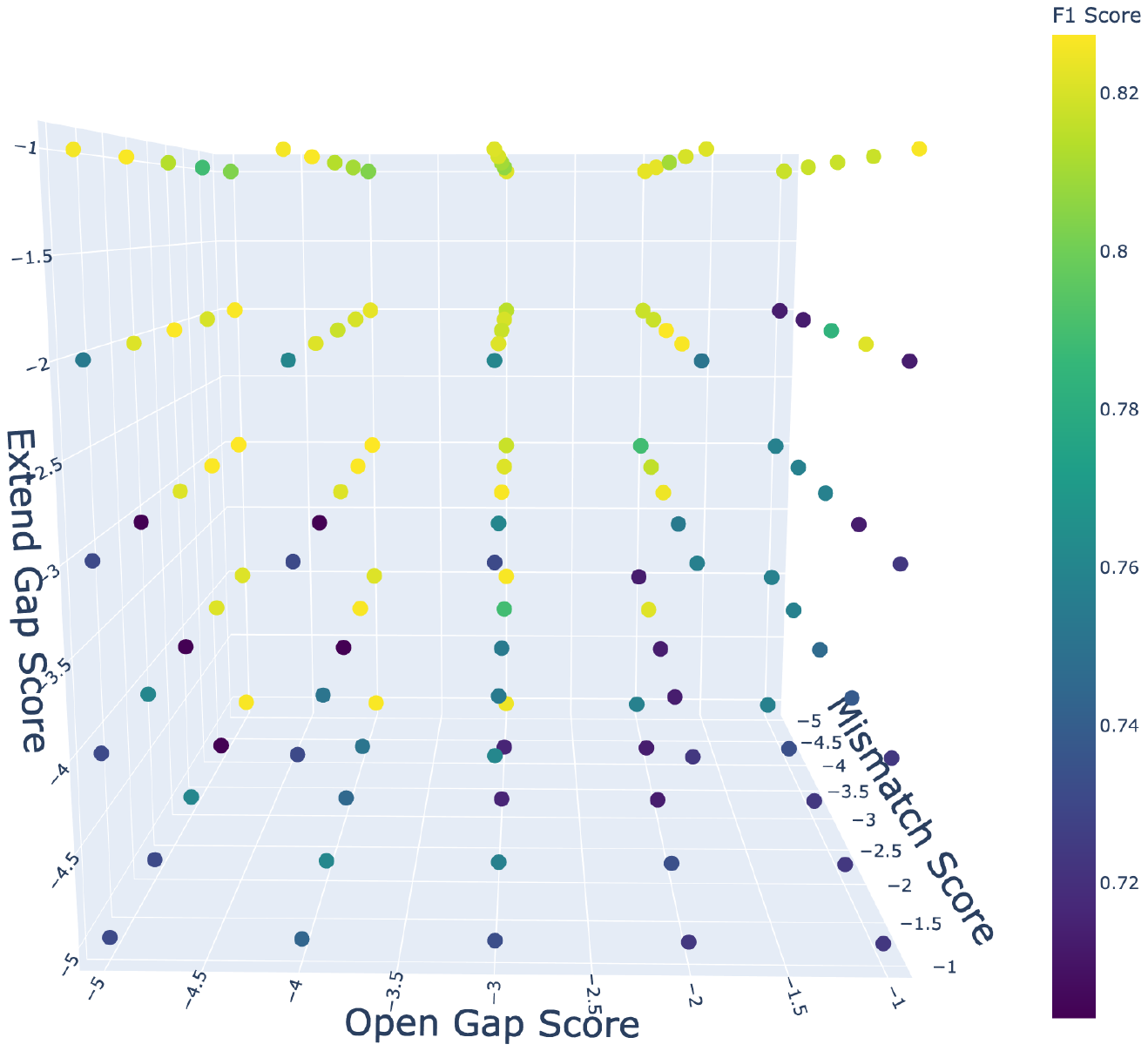}
                \subcaption{\textit{fox-8}}
            \end{subfigure}
            \caption{Evaluation of mismatch/open gap/extend gap scores\label{fig:scores_scatter}}    
        \end{figure*}

        In bio-informatics, alignments 
        can indicate whether sequences share a common evolutionary origin or have similar structural functions \cite{NEEDLEMAN1970443}. There are two primary approaches to aligning sequences: global and local. For our purposes, we adopt the global alignment approach: unlike local alignment, which focuses on isolated regions of similarity and may miss important differences in the rest of the sequences, global alignment ensures that these differences are accounted for, providing a more comprehensive basis for our comparative analysis. 
        Global pairwise alignment uses four key parameters: \textit{match}, \textit{mismatch}, \textit{open gap}, and \textit{extended gap}, as introduced in \cite{NEEDLEMAN1970443} and \cite{WATERMAN1976367}. A \textit{match} indicates the presence of identical nucleotides at the same index, a \textit{mismatch} indicates their disparity, an \textit{open gap} indicates the initial insertion or deletion and an \textit{extended gap} reflects the extension of an existing gap. 
        Initially, we experimented by setting the initial scoring values as follows: match = 0, mismatch = -1, open gap = -1 and extended gap = -1. This ensures that the alignment returns a score \eqref{eq:score} equivalent to the Levenshtein distance (indeed penalties for mismatches correspond to the cost of substitutions in the Levenshtein distance, and penalties for gaps align with the costs of insertions and deletions). We then conducted several experiments in which we maintain the structure of the Levenshtein match score (=0) while systematically varying the penalty scores for mismatches, open gaps, and extended gaps. 
        We examined over 300 combinations of scoring penalties, \autoref{fig:scores_scatter} illustrates a broad spectrum of them  ranging from -1 to -5 (beyond a penalty of -5 we observed a reduction in the  classification performance).
        The optimal scoring system identified from these tests is 0 for matches, -5 for mismatches, -4 for open gaps, and -5 for extended gaps. This system achieves the highest F1 score in all five cases tested ($\alpha$ in \autoref{tab:results}). Furthermore, the results also show a slight improvement over the results obtained with Levenshtein distance metric ($\lambda$ in \autoref{tab:results}). Based on these findings, we will proceed with the sequence alignment metric and using the optimal scoring configuration: match = 0, mismatch = -5, open gap = -4, and extended gap = -5.

        \input{latex_template/tables/merged_table}

        As discussed in \cite{NEEDLEMAN1970443}, the optimal alignment is the highest achievable score value, where the score is calculated as follows:
\small{
    \begin{align}\label{eq:score}
        \mbox{score} = & \text{ } \#\textit{matches} \times \textit{match\_score} + \#\textit{mismatches} \times \textit{mismatch\_score} \notag \\ 
                       & + \#\textit{open\_gaps} \times \textit{open\_gap\_score} \notag \\ 
                       & + \#\textit{extended\_gaps} \times \textit{extended\_gap\_score}
    \end{align}
}

\begin{figure}[tbh]
            \centering
            \includegraphics[width=.7\linewidth]{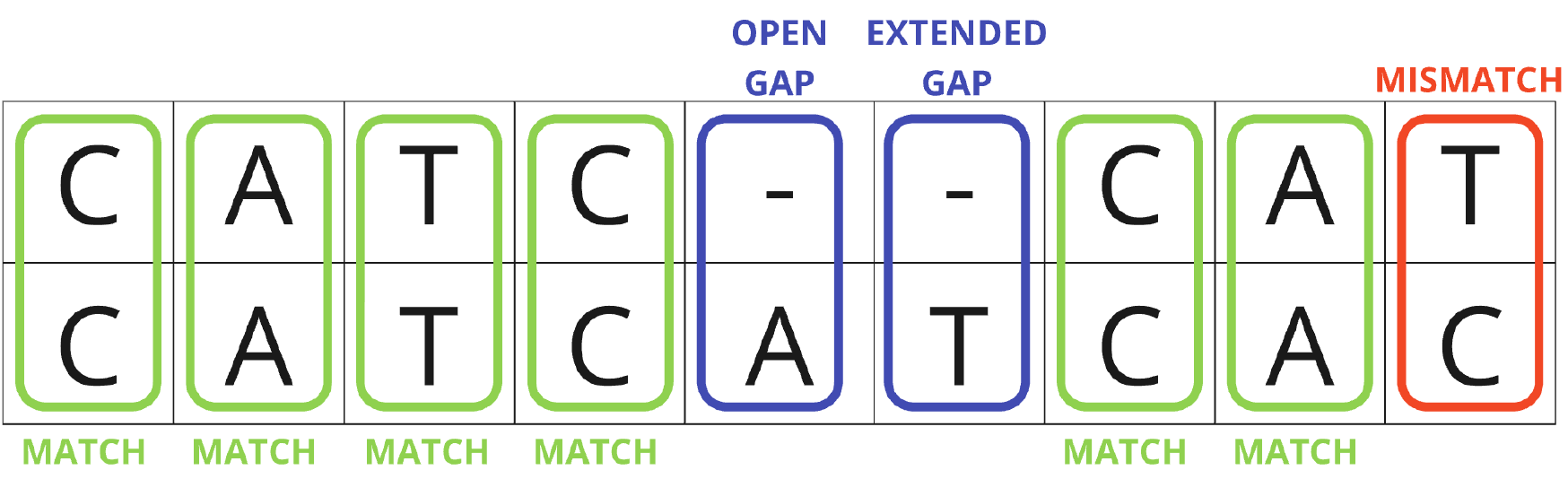}
            \caption{An example of optimal alignment between two sequences} \label{fig:scoring}
        \end{figure}

        \normalsize
        Since we assign the match score a value of 0, the overall score can assume values $\in [|aligned\_seq| \times \text{min}(\text{mismatch\_score}, $ \newline $\text{open\_gap\_score},  \text{extend\_gap\_score}) , 0]$.
        Let the reader consider the sequences $A =$ \verb|CATCCAT| and $B =$ \verb|CATCATCAC|. One of the best optimal alignments (there may be multiple optimal alignments) for the two sequences is the one in \autoref{fig:scoring}.

        After aligning two sequences, the similarity score $\mathbb S$ between them is calculated normalizing \eqref{eq:score} with the Min-Max data normalization method, widely used in the literature \cite{JAIN20052270}:
        \begin{align}\label{eq:sim}
            \mathbb S = \frac{\mbox{score} - \mbox{min\_possible\_score}}{\mbox{max\_possible\_score} - \mbox{min\_possible\_score}}
        \end{align}
        
        Now, the similarity score $\mathbb S$ varies between 0, when the two aligned sequences differ for every nucleotide, and 1, when the two sequences are identical. The similarity value between the sequences in \autoref{fig:scoring} is $\frac{(6 \times 0 + 1 \times (-5) + 1 \times (-4) + 1 \times (-5)) - (-45)}{(0 - (-45))} = 0.69$, since $\mbox{min\_possible\_score} = |aligned\_seq| \times \text{min}(\text{mismatch\_score}, $ \newline$\text{open\_gap\_score},  \text{extend\_gap\_score}) = 9 \times \text{min}(-5, -4, -5) = -45$.

    \subsubsection{WeightedLCS}
        In addition to the score obtained from the pairwise sequence alignment procedure, we will also use a new metric we have coined, to rank accounts of unlabeled species: 
        \begin{equation}\label{eq:weighted_LCS}
            \textit{weightedLCS}_s = \frac{|LCS_s|}{\frac{1}{|s|} \sum_{i=1}^{|s|} |\text{DNA}_{i}|} \times |s|
        \end{equation}    
        where $|$DNA$_i$$|$ is the length of the digital DNA of user $i$ in species $s$. The formula considers both the relative similarity among DNA sequences within a species and the size of the population contributing to that similarity. Typically, the \textit{weightedLCS} value for a species consisting mostly of genuine users ranges from 0 to 2. This range has been observed in experiments conducted on datasets of genuine users, where the LCS tends to be considerably shorter compared to the average DNA length.

    \subsubsection{Classification of unlabeled species.}\label{sec:unlabeled}

        To classify the unlabeled species (those colored gray in \autoref{fig:classificationProcess}), we use an iterative procedure that leverages both the similarity score $\mathbb S$ (\eqref{eq:sim}) and the \textit{weightedLCS} (\eqref{eq:weighted_LCS}). The overall idea is to compare the sequences of the LCSs of the unclassified species with those of \textit{gGenuine} \textit{gSpamBot}. The visual representation in \autoref{fig:histo_images} offers a detailed insight into the classification procedure, while Algorithm~\ref{alg:pseudocode} details the corresponding pseudocode.

        \input{latex_template/paper/methodology/pseudocode}

        The procedure is as follows:

        \input{latex_template/paper/methodology/gray_classification}

    \begin{enumerate}
        \item A species $s$ is included in \textit{gGenuine} if it meets one of the following criteria:
            \begin{enumerate}[label=(\roman*)]
                \item it has a similarity score to \textit{gSpamBot} that is lower than the similarity score between \textit{gGenuine} and \textit{gSpamBot}. This criterion allows us to reasonably infer that the species is likely to contain genuine users.
                Specifically, the similarity score $\mathbb S$ between $s$ and \textit{gSpamBot} is less than, or equal to, the similarity score $\mathbb{M}$ between \textit{gGenuine} and \textit{gSpamBot} $\And$  \textit{weightedLCS}$_s$ is less than, or equal to, $|s| \times 0.9$.\footnote{The second condition is used to avoid classifying in \textit{gGenuine} a species with a very high ratio, thus close to $1$, between its LCS and the average length of DNA sequences. This would be an indicator of bot behavior patterns.}
                
                \item \textit{weightedLCS}$_s$ is less than $2$. 
            \end{enumerate}

        \item A species $s$ is included in \textit{gSpamBot} if it meets one of the following criteria: 
            \begin{enumerate}[label=(\roman*)]
                \item it has a level of similarity to \textit{gSpamBot} that exceeds a certain threshold designed to account for the inherent differences between bots and genuine accounts.
                Specifically, the similarity $\mathbb S$ between $s$ and \textit{gSpamBot} is  \( \geq \mathbb{L}\), where \(\mathbb L := 1 - \left(\frac{1 - \mathbb{M}}{x}\right) \), \( x \geq 1 \).

                
                \item  \textit{weightedLCS}$_s$ $>$ 4. A gap between 2 and 4 is deliberately left to accommodate species whose \textit{weightedLCS} values are neither too low nor too high, thus postponing their classification.
    
            \end{enumerate}
        
        \item If there are species remaining unclassified, thus falling within the similarity range $(\mathbb{M}, \mathbb L)$ and with \textit{weightedLCS} values between 2 and 4, they are merged into a single large group. This group is then given as input to the LCS clustering algorithm (described in \autoref{subsec:initial_clustering}) to form new "gray" species. The classification process restarts from step 1\footnote{To ensure algorithmic completeness, in edge cases where the LCS clustering algorithm returns the same species divisions as the input, the classification would fall on the \textit{gGenuine} group, since no discernible bot characteristics are observed.}.
    \end{enumerate}


    \subsubsection{The importance of \textit{weightedLCS}} 
        The inclusion of the \textit{weightedLCS} metric is crucial because it accounts for scenarios in which species may appear to have characteristics not so dissimilar to \textit{gSpamBot}, yet have a low \textit{weightedLCS}, indicating genuine characteristics. Conversely, species seemingly dissimilar to \textit{gSpamBot} may have high \textit{weightedLCS} values, suggesting bot-like characteristics despite low DNA similarity score. 
        To better understand the role that \textit{weightedLCS}  plays in the classification process, let the reader consider a \textit{gSpamBot} group whose LCS is composed entirely of nucleotides \verb|T|. 
        Then, imagine an unknown, highly populated species with 1) a high LCS, comparable in length to the average DNA sequences within that species, and 2) composed entirely of nucleotides \verb|A|. 
        Obviously, the similarity score $\mathbb S$ between \textit{gSpamBot} and the species under analysis will be very low, which would tend to rule out that the species accounts are bots.
         Instead, if we also consider the value of \textit{weightedLCS}, we can well classify those accounts that have bot-like characteristics, but appear to be genuine by DNA similarity.

    \subsubsection{Estimation of parameter $x$}
        As defined in \autoref{sec:unlabeled}, a condition for a species to be included in the spambot group is that the similarity score $\mathbb{S}$ between the two species is greater than a certain limit $\mathbb L$, which depends on the similarity $\mathbb M$ between spambots and genuine accounts and a parameter $x$ $\ge$ 1.  As $x$ increases, so does $\mathbb L$, and so does the number of accounts that are mistaken for bots. Conversely, as $x$ decreases, $\mathbb L$ decreases, and it is possible that real accounts are misclassified as bots. 
        \autoref{tab:x_eval} shows that $2$ is the most consistent value of $x$ to avoid misclassification, for different combinations of \textit{Cresci-17} sub-sets and for \textit{fox-8}. 

        \input{latex_template/tables/Xtable}

%% file: latex_template/tables/merged_table.tex
\begin{table}[tbh]
\caption{Our results using $\lambda$ (Levenshtein Distance) \text{vs} $\alpha$ (Sequence Alignment: match=0, mismatch=-5, open\_gap=-4, extend\_gap=-5)}
\label{tab:results}
\begin{tabular}{clccccccc}
\toprule
\multicolumn{1}{l}{} &  & \multicolumn{3}{c}{$\lambda$} &  & \multicolumn{3}{c}{$\alpha$}    \\ \cline{3-5} \cline{7-9} 
\multicolumn{1}{l}{} &  & ACC      & F1      & MCC      &  & ACC         & F1          & MCC \\ \cline{3-5} \cline{7-9} 
Datasets             &  &          &         &          &  &             &             &     \\ \cline{1-1}
gen $\cup$ ss1       &  & 99       & 97      & 96       &  & 99          & 97          & 96  \\
gen $\cup$ ss2       &  & 94       & 94      & 89       &  & \textbf{95} & \textbf{95} & 89  \\
gen $\cup$ ss3       &  & 99       & 96      & 96       &  & 99          & 96          & 96  \\
Cresci-17            &  & 95       & 95      & 89       &  & 95          & \textbf{96} & 89  \\
fox-8                &  & 80       & 83      & 63       &  & 80          & 83          & 63  \\ \bottomrule
\end{tabular}
\end{table}

%% file: latex_template/paper/methodology/pseudocode.tex
\begin{algorithm}[tbh]
\caption{Classification of unlabeled species}\label{alg:pseudocode}
\begin{algorithmic}[1]
    \Procedure{classify\_via\_similarity}{$gGen: Group$, $gSpam: Group$, $species\_left: list[Group]$, $x: int$}
    \vspace{0.3em}
    \State $M \gets $ \Call{dna\_similarity}{gGen.lcs, gSpam.lcs}
    \State $gSpamLCS \gets  gSpam.lcs$
    \vspace{0.3em}
    
    \For{each $species$ in $species\_left$}
        \vspace{0.3em}
        \State $gSpam\_sim \gets $ \Call{dna\_similarity}{species.lcs,gSpamLCS}
        \State $weigthedLCS \gets $ \Call{weigthedLCS}{species}
        \State $limit \gets 1 - ((1-M)/x)$
        \vspace{0.3em}
        \State \# Case (1)
        \If{($gSpam\_sim \leq M$ and $weigthedLCS \leq (\textsc{len}(species.users) \times 0.9)$) or $weigthedLCS < 2$}
            \State $gGen.add(species.users)$
        \vspace{0.3em}
        \State \hspace{-1.7em} \# Case (2)
        \ElsIf{$gSpam\_sim \geq limit$ or $weigthedLCS > 4$}
            \State $gSpam.add(species.users)$
        \EndIf    
    \EndFor

    \vspace{0.5em}
    \State \# remove from $species\_left$ the classified species
    \vspace{0.3em}
    
    \State \# Case (3)
    \If{$species\_left$ is not empty}
        \State $re\_grouped \gets \text{LCS\_re\_clustering}(species\_left)$
        \State \Return \Call{classify\_via\_similarity}{$gGen$, $gSpam$, $re\_grouped$, $x$}
    \EndIf

    \EndProcedure

\end{algorithmic}
\end{algorithm}

%% file: latex_template/paper/methodology/gray_classification.tex
\begin{figure*}[htb]
    \begin{minipage}{.35\textwidth}
        \begin{subfigure}{\textwidth}
            \includegraphics[width=\textwidth]{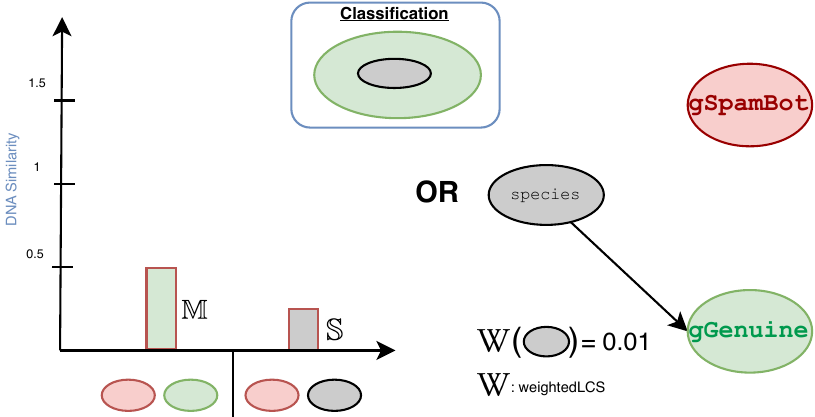}
            \subcaption{Case 1. (i) or (ii)}
        \end{subfigure}
    \end{minipage}
    \hfill
    \begin{minipage}{.35\textwidth}
        \begin{subfigure}{\textwidth}
            \includegraphics[width=\textwidth]{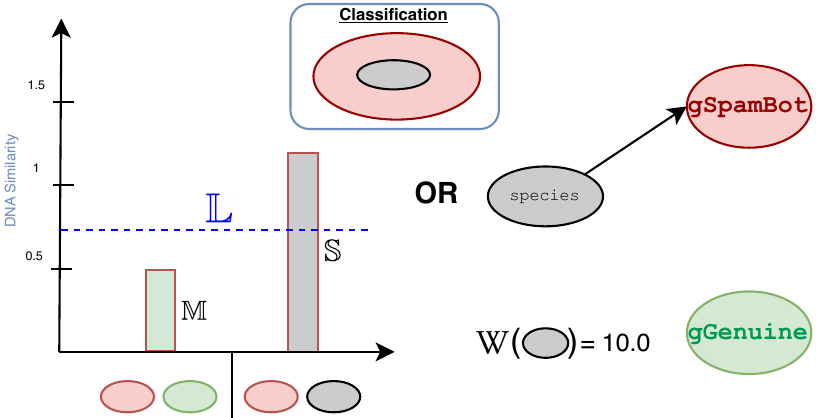}
            \subcaption{Case 2. (i) or (ii)}
        \end{subfigure}
    \end{minipage}
    \hfill
    \begin{minipage}{.18\textwidth}
        \begin{subfigure}{\textwidth}
            \includegraphics[width=\textwidth]{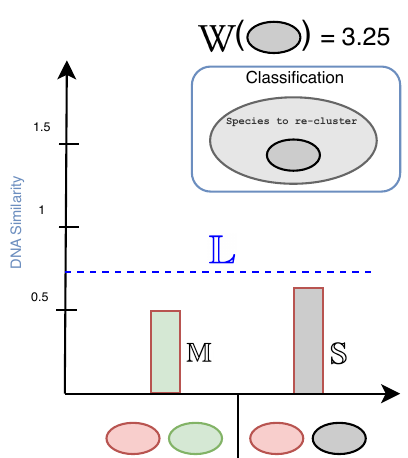}
            \subcaption{Case 3.}
        \end{subfigure}
    \end{minipage}
    \caption{Unlabeled species classification\label{fig:histo_images}}
\end{figure*}

    

%% file: latex_template/tables/Xtable.tex
\begin{table*}[htb]
\centering
\caption{Evaluation of parameter $x$}\label{tab:x_eval}
\begin{tabular}{lllccclccclccclccclccc}
\toprule
 &     &  & \multicolumn{3}{c}{\textbf{gen $\cup$ ss1}} &  & \multicolumn{3}{c}{\textbf{gen $\cup$ ss2}} &  & \multicolumn{3}{c}{\textbf{gen $\cup$ ss3}} &  & \multicolumn{3}{c}{\textbf{Cresci-17}} &  & \multicolumn{3}{c}{\textbf{fox-8}} \\ \cline{4-6} \cline{8-10} \cline{12-14} \cline{16-18} \cline{20-22} 
 & $x$ &  & ACC                & FP         & FN        &  & ACC                & FP         & FN        &  & ACC                & FP         & FN        &  & ACC              & FP       & FN       &  & ACC                  & FP   & FN   \\ \cline{2-2} \cline{4-6} \cline{8-10} \cline{12-14} \cline{16-18} \cline{20-22} 
 & 1   &  & 90                 & 423        & 9         &  & 94                 & 260        & 141       &  & 88                 & 470        & 3         &  & 95               & 288      & 157      &  & 71                   & 617  & 35   \\
 & 2   &  & \textbf{99}        & 31         & 25        &  & \textbf{95}        & 236        & 141       &  & \textbf{99}        & 8          & 28        &  & \textbf{95}      & 249      & 183      &  & \textbf{80(F1=83)}   & 387  & 64   \\
 & 3   &  & 99                 & 13         & 33        &  & 95                 & 236        & 141       &  & 99                 & 8          & 28        &  & 95               & 249      & 183      &  & 80(F1=81)            & 303  & 150  \\
 & 4   &  & 99                 & 13         & 33        &  & 95                 & 236        & 141       &  & 99                 & 8          & 28        &  & 95               & 249      & 183      &  & 78                   & 451  & 49   \\ \bottomrule
\end{tabular}
\end{table*}

%% file: latex_template/paper/evaluation_results/results.tex
   


    Classification performance was evaluated using well-established standard metrics, including accuracy, F1 (the harmonic mean of precision and recall), and the Matthews Correlation Coefficient (MCC), which measures the correlation between the predicted and actual classes of the samples. 
    
    During the testing phase, \textbf{gen} from \textit{Cresci-17} (presented in \autoref{subsec:c17}) were alternately paired with \textbf{ss1}, \textbf{ss2}, and \textbf{ss3}, and we have also considered the entire dataset. This approach was chosen to comprehensively evaluate the classifier's performance across different types of spambots, and to ensure its robustness in distinguishing genuine accounts from various spambot behaviors and characteristics.
    We also included in our tests one of the most recent publicly available datasets, \textit{fox-8}. A different alphabet than $\mathbb{B}^{3}_{\textit{type}}$ was used for the analysis of \textit{fox-8}. For this study, tweets are coded according to their content (plain tweet, contains url, contains hashtags).

    For the \textit{fox-8} dataset, we have the weakest results, which is to be expected since this is a newly discovered and highly evolved family of bots that use LLMs to produce the content to be posted. To the best of our knowledge, the best result obtained in the literature on this dataset consists of an F1 = 0.84 (versus our 0.83 shown in \autoref{tab:results}) using the OpenAI tool trained to recognize LLM-generated text\footnote{New AI classifier for indicating AI-written text: \url{https://openai.com/blog/new-ai-classifier-for-indicating-ai-written-text}}\cite{fox8}. 
    Since our proposal is based on a rule-based system without the use of ML/DL models, it provides transparency and direct  interpretability, as shown in \autoref{fig:classificationProcess} and detailed in Algorithm~\ref{alg:pseudocode}. 
    This makes the results easy to understand and explain.


    \subsection{Further considerations on \textit{fox-8}\label{subsec:considerations}}

    \begin{figure}[tbh]
        \centering
        \includegraphics[width=.9\linewidth]{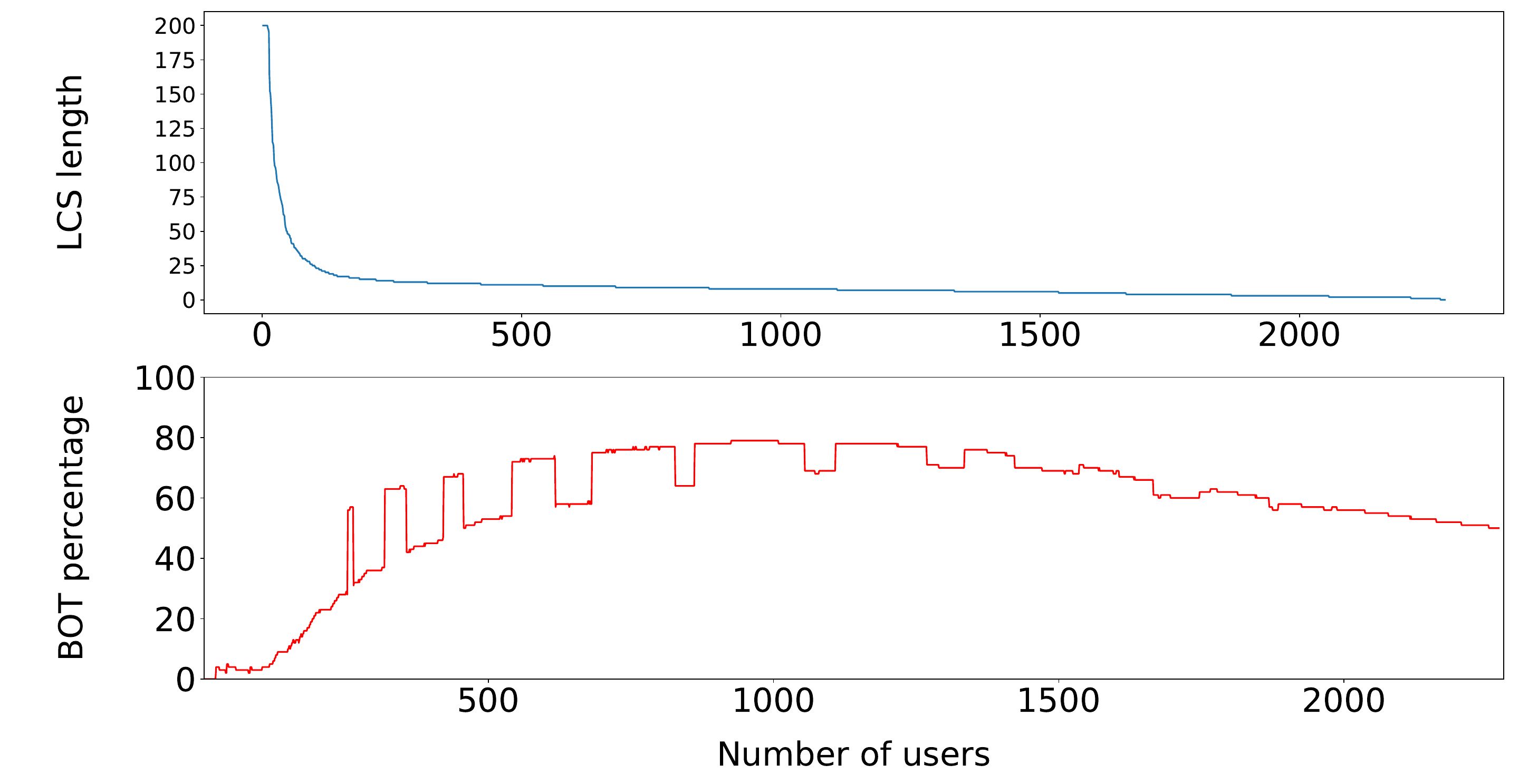}
        \caption{Anomaly in \textit{fox-8} vector $\mathbb V$} \label{fig:fox8_anomaly}
    \end{figure}

\begin{figure}[h!]
        \centering
        \includegraphics[width=.9\linewidth]{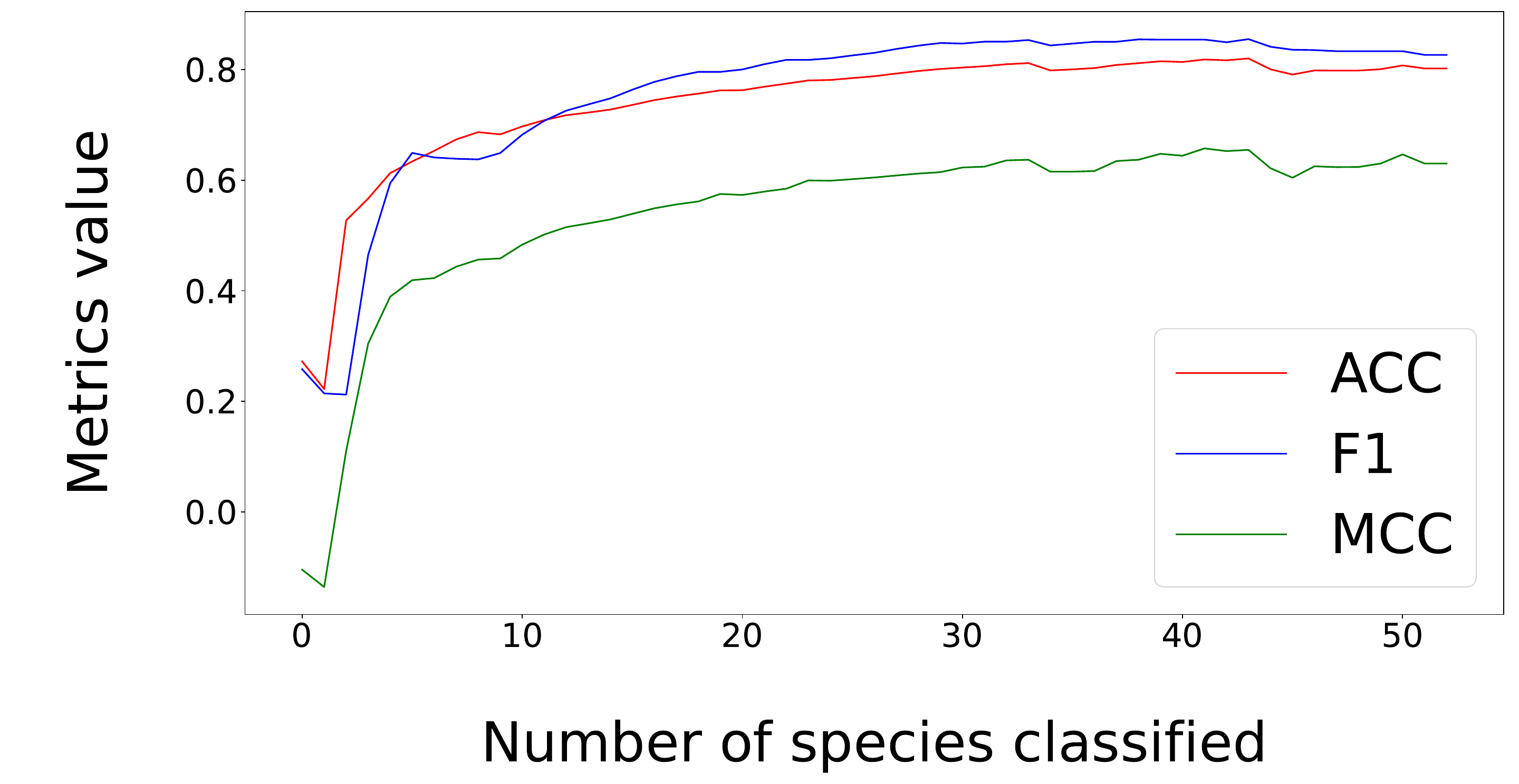}
        \caption{\textit{fox-8} Classification Trend}\label{fig:trendfox8}
    \end{figure}

    A peculiarity of the \textit{fox-8} dataset can be seen in \autoref{fig:fox8_anomaly}. 
    At the top (blue) is the LCS curve, at the bottom (red) is the percentage of bots detected for each group of k users defined at the beginning of \autoref{subsec:initial_clustering}.
    In the blue curve, each x abscissa represents a group of x users, while the y value indicates the length of the longest LCS found among all possible groups of x users in the dataset.
    The red curve represents the same groups, but for each it shows the percentage of bot users within that group.
    In the particular case of \textit{fox-8}, the early part of the red curve shows that the percentage of bots is much smaller than the percentage of real users. This contradicts the common intuition that groups with a long LCS are bot accounts.
    In fact, we found that with k between about 2 and 250, the groups consist mostly of real accounts.
    The bad implication is that our initial \textit{gSpamBot} does indeed contain many genuine accounts. This happens because this group has a very long LCS, formed by the U character (which encodes the action: tweet containing a URL). 
    The other side of the coin, however, shows us something positive: our procedure is able to correct the error, as less clear types are associated with \textit{gSpamBot} or \textit{gGenuine}. This can be seen in \autoref{fig:trendfox8}, which demonstrates how classification performance increases as the procedure progresses.
    In this case, trying to code the account behavior with a more detailed alphabet led to excellent classification results.

%% file: latex_template/paper/related_work.tex
A variety of methods have been used in the literature to detect social bots, and over the years, more and more up-to-date techniques have been applied to cope with the evolving nature of these types of accounts. 
The following excursus is not meant to be exhaustive but rather to give an idea of the possible dimensions of this type of task.
Feature-based approaches are the most commonly proposed. Enhancing datasets with additional features can be valuable for detecting complex behaviors, as shown by Cozza et al. \cite{cozza_hotel}. Similarly, Yang et al.~\cite{yang2013empirical} identified features that effectively detect Twitter spammers, demonstrating good performance on two custom datasets. Botometer~\cite{yang2022botometer} is a bot-detection tool that uses supervised machine learning on over $1000$ features. However, as recent studies reported~\cite{fox8}, Botometer may fail to classify bot accounts using LLMs. Efthimion et al.~\cite{efthimion2018supervised} proposed an approach based on features like username length, temporal patterns, followers/friends ratio, sentiment expression, and reposting rate for bot classification. Hayawi et al.~\cite{hayawi2022deeprobot} developed \emph{DeepProBot}, a deep-neural network that exploits user profile data and an LSTM network, achieving an AUC of $0.72$ on the Cresci-rtbust dataset~\cite{mazza2019rtbust}. 
Kudugunta et al.~\cite{kudugunta2018deep} also proposed an LSTM architecture based on tweet content and user metadata, achieving an AUC $> 96\%$ on the \textit{Cresci-17}. 
Feng et al.~\cite{feng2021botrgcn} developed \emph{BotRGCN}, a Graph Neural Networks (GNNs) architecture that exploits user semantic information from tweets. The literature includes additional approaches that leverage the concept of digital DNA, a type of account behavior modeling that is the inspiration for this work. For example, Pasricha et al.~\cite{pasricha2019detecting} combined digital DNA with lossless compression algorithms and used logistic regression for classification. Gilmary et al.~\cite{gilmary2023entropy} developed a framework based on digital DNA that uses sequence entropy for account classification. Di Paolo et al.~\cite{dipaolo2023dna} transformed DNA sequences into images and then used convolutional neural networks (CNNs) to classify the images. Chawla et al.~\cite{chawla2023hybrid} used BERT~\cite{DBLP:journals/corr/abs-1810-04805} to extract sentiment information (positive, negative, or neutral) from tweets and proposed a new alphabet for digital DNA.

The aforementioned works (\cite{pasricha2019detecting,dipaolo2023dna,chawla2023hybrid,gilmary2023entropy}) use digital DNA to encode account behavior. Once the DNA string is obtained, it is processed to obtain new features, which in turn are used to classify the accounts. 
There are at least two relevant differences between the cited work and ours. 
First, our work relies exclusively on operations on the native DNA string. By playing with the degree of similarity between strings, and not just equality as in previous work, we cleanly manage to detect as bot or not accounts that are in limbo. To the best of our knowledge, this is the first approach that extends the original idea of modeling account behavior through digital DNA and detecting its nature using the notion of LCSs (\cite{DBLP:journals/expert/CresciPPST16,dna_pres}) by considering not only \textit{identical} strings, but also \textit{similar} ones.
In this way, we can draw on decades of bioinformatics research and make anomalous accounts emerge using standard DNA sequence alignment tools. 
Second, our proposal does not use machine learning techniques. 
We are aware that this choice goes against the current research trend, but it seems appropriate.  The approach is that of a rule-based, well-defined system (see \autoref{fig:classificationProcess}, \autoref{fig:histo_images} and Pseudocode \ref{alg:pseudocode}) and the detection performance are very promising.

%% file: latex_template/paper/conclusion.tex
This study presents an innovative method that uses DNA similarity algorithms to efficiently classify social media users, 
providing a flexible solution to combat the ever-evolving strategies of social bots.
The classification process begins by grouping users into species based on the degree of similarity in user behavior. Those with bot-like characteristics are then labeled as such, and the same is done for authentic users. The novelty of this work, compared to previous work that also relies on the analysis of online behavior to detect anomalous accounts, is that through an iterative process, it is possible to classify accounts with less distinct characteristics. The result is achieved by applying standard sequence similarity algorithms along with a new metric coined for the occasion that manages to correct detection where similarity is insufficient. Ours is a largely intuitive, rule-based system, which can be advantageous in terms of interpretability of the result compared to more widely used ML/DL-based methods (which it is not our intention to fault). Our future focus is to improve the species clustering process, recognizing that establishing high quality species will improve classification accuracy. In addition, we plan to adapt the presented procedure to a \emph{multi-class} classification, aiming to identify different social bot breeds from the clustered species.